%% file: main.tex
\documentclass[10pt, conference, letterpaper]{IEEEtran}
\usepackage{url}
\usepackage{multirow}
\usepackage{cite}
\usepackage{amsmath,amssymb,amsfonts}
\usepackage{algorithmic}
\usepackage{graphicx}
\usepackage{textcomp}
\usepackage{xcolor}
\def\BibTeX{{\rm B\kern-.05em{\sc i\kern-.025em b}\kern-.08em
    T\kern-.1667em\lower.7ex\hbox{E}\kern-.125emX}}

\newcommand{\eg}{e.g.,\ }
\newcommand{\ie}{i.e.,\ }

\newcommand{\fig}{Figure\ }

\begin{document}

\title{Watching the Watchers: Nonce-based Inverse 
Surveillance to Remotely Detect Monitoring}

\author{
    \IEEEauthorblockN{Laura M. Roberts\IEEEauthorrefmark{1}\IEEEauthorrefmark{2}, David Plonka\IEEEauthorrefmark{1}}
    \IEEEauthorblockA{\IEEEauthorrefmark{1}Akamai Technologies}
    \IEEEauthorblockA{\IEEEauthorrefmark{2}Princeton University}
}

\maketitle
\input{abstract}


\input{introduction}
\input{bgrelatedwork}
\input{method}
\input{data}
\input{results}
\input{contributions}

\pagebreak
\section*{Acknowledgments}
\input{acknowledgment}

\bibliographystyle{IEEEtran}
\bibliography{IEEEabrv,references}

\end{document}

%% file: abstract.tex
\begin{abstract}

Internet users and service providers do not often know when traffic
is being watched but desire a way to determine when, where,
and by whom.  We present NOISE, the Nonce Observatory for
Inverse Surveillance of Eavesdroppers, a method and system that
detects monitoring by
disseminating nonces---unique, pseudorandom values---in traffic
and seeing if they are acted upon unexpectedly, indicating that the
nonce-laden traffic is being monitored. Specifically, we embed 64-bit
nonces innocuously into IPv6 addresses and {\em disseminate} these 
nonces Internet-wide using a modified traceroute-like tool that makes 
each outbound probe's source address unique.  We continually monitor for
subsequent nonce {\em propagation}, \ie activity or interest 
involving these nonces, \eg via packet capture on our system's
infrastructure.  Across three experiments and four months,
NOISE detects monitoring more than 200k times,
ostensibly in 268 networks, for probes destined for 437 networks.
Our results reveal:
{\em (a)} data collection for security incident handling, 
{\em (b)} traffic information being shared with third parties, 
and 
{\em (c)} eavesdropping in or near a large commercial peering exchange.

\end{abstract}

%% file: introduction.tex
\section{Introduction}

Internet users and services exchange content worldwide every day.
This traffic
traverses routers and exchange points far and wide,
but neither those users nor service providers typically know who, if anyone, is
watching that traffic.
In today's Internet, the community has deemed
pervasive monitoring to be a threat~\cite{rfc7258}.
Knowledge of such monitoring is of significant interest because surveillance:
$(a)$ can threaten quality of service, 
\eg when surveillance aids reconnaissance prior to
intrusions, thefts of data, or denial-of-service (DoS) attacks;
and $(b)$ can threaten the privacy of end-users,
risking the reputations of users and service providers
when private information is exposed.
Thus, the goal of our work is to detect traffic monitoring,
Internet-wide, detecting monitoring organizations 
and monitoring systems, \eg network firewalls, email filters, and even wiretaps.
We also want to know where they are,
be it on network links or edges, and to classify such
systems when they are of a common type. Furthermore, we want to
detect subsequent data sharing, \eg
when information about traffic is shared with third parties,
because this exacerbates challenges to privacy.

Discovery and disclosure of Internet monitoring by nation
states~\cite{demchak2018china} and
other institutions~\cite{farsightPassiveDNS,renIsacPdns} have alerted the
community to the presence of such surveillance, and 
one might expect Internet surveillance not be
hard to find by those willing to look.
In this work we aim to answer the question, 
``Can we build a system that remotely detects monitoring?''
To that end, we introduce the Nonce Observatory for Inverse Surveillance of Eavesdroppers (NOISE), 
a method and system to 
detect monitoring by disseminating nonces, which are single-use, 
pseudorandom values, 
 and stealthily
listening for their subsequent propagation.

First, we {\em actively} {\em disseminate} nonces, 
\ie we 
initially transmit them as identifiers in Internet traffic 
(\eg 
as a packet's IPv6 source address in an active measurement survey),
and then we {\em passively}
listen for a surveillant to {\em propagate} or convey
a nonce to somewhere unusual, \eg to retransmit
it in a response packet 
or use it to form a reverse DNS
query.  Because the nonces are unique, 
we are able to
correlate their dissemination with any subsequent propagations.
And because we disseminate nonces
in hop-limited packets via an enhanced traceroute, we glean topological information on 
paths that nonces traverse, helping to locate surveillants when detected.
Although the technique is not IPv6-specific, our current
system detects monitoring of only IPv6 traffic with the
expectation that anyone who
monitors IPv6 almost certainly monitors IPv4 as well.
NOISE reports monitoring without regard to legitimacy or intent, \eg
monitoring that represents security best practices at or near hosts 
and also {\em eavesdropping}, \ie monitoring in the middle of the communication path.  

\looseness=-1
Our contributions comprise both methods and results:
{\em (i)} a practical inverse surveillance method to detect Internet traffic monitors;
{\em (ii)} a modified traceroute that can show when probes propagate further than shown by traditional traceroutes;
{\em (iii)} detection of traffic monitoring in $268$ networks;
{\em (iv)} detection of eavesdropping in a commercial Internet exchange;
{\em (v)} detection of networks sharing traffic information with third parties, \eg public DNS services;
{\em (vi)} detection of automated security practices including: forensic DNS queries, logging and retention of traffic records, and aggressive counter-probes; and
{\em (vii)} results validation using ground truth from interview with experts on three networks where monitoring was detected.

%% file: bgrelatedwork.tex
\section{Background and Related Work}
\label{sec:bgrelated}

Our system implements a form of {\em inverse surveillance}~\cite{mann2003sousveillance} 
but focuses on the detection and
monitoring of surveillants in the Internet rather than in the ``real world.''
In contrast, Mann coined
the term ``sousveillance''~\cite{mann2004sousveillance} which focuses on using
technology such as wearable devices to enable data collection,
\eg citizens recording video of surveillants.

In our inverse surveillance of Internet traffic, we watch for both interception
that is lawful (LI) and potentially unlawful, {\em i.e.,} regardless of purpose,
whether it is {\em (a)} likely innocuous, such as current best 
practice monitoring of one's own network for performance or
security, or {\em (b)} potentially nefarious, such as a malicious
party or nation state surreptitiously monitoring traffic at a host
or within an IXP.
While much prior engineering and research
work has to do with surveillance in networks, we know of
few that focus on surveilling the surveillants~\cite{stoll1990cuckoo}.

Prior works~\cite{RFC3972,dnsSpiesCox,McRae2007PhightingTP,honeytokenWikipedia,honeytokens} inform our nonce-based detection, 
\eg nonces used in DNS labels or ``honeytokens'' in data objects.
There are prior works that relate to ours in a number of areas.

First, our work is not the first to remotely detect traffic monitoring.
Some prior works~\cite{vandersloot2018quack,mcdonald2018403} detect
types of censorship that entailed monitoring. Instead, we develop a
technique having detection of surveillance as its primary goal.

Second, because our method requires dissemination of nonce-laden identifiers,
we leverage existing means to do so. Given that locating
surveillance points is also our goal, we choose to
augment yarrp~\cite{imc16yarrp,Beverly:2018:IBS:3278532.3278559},
a high-performance traceroute-like tool, thus simultaneously discovering
the topology in which surveillance is taking place, if and when detected.
In this, we are also inspired by
TCP Sidecar~\cite{sherwood2006touring} to piggyback new measurements
atop existing traffic.

Third, our method detects the propagation of nonce-laden
identifiers or nonces themselves and identifies candidate networks
that observe and propagate them.  In this way,
it bears some similarities to efforts in validating and verifying
network paths~\cite{bu2018s} or routes~\cite{wong2007truth}. Given
that practical path validation does exist today, we instead explore
whether it is possible to detect unexpected divergence of
traffic from its expected path, such as an 
eavesdropper passing information to a third party that is neither
the source nor intended destination of the associated traffic.
As in the evaluation of Alibi Routing~\cite{levin2015alibi}, we
attempt to geolocate~\cite{edgescape} where our traffic, or information about that
traffic, may have unexpectedly traveled.

\looseness=-1
Last, because our detection of surveillance depends on
opportunity to witness suspicious actions of surveillants
that imply their having observed our traffic, our method
is somewhat inspired by the notion of opportunistic
measurements~\cite{casado2005opportunistic}.

%% file: method.tex
\section{Method}
\label{sec:method}
Aiming to remotely detect monitoring,
we realize a system having two primary operations:
{\em (a)} {\em active} conveyance or
dissemination of nonces to distant destinations by
placing them in transport header fields of traffic we transmit in periodic
measurement campaigns and {\em (b)} {\em passive} observation,
listening for reactions to that nonce-laden traffic that propagate
back to our system's inverse monitoring components.

\subsubsection{\bf Active components}
NOISE disseminates nonces via traceroute-like surveys, 
essentially masquerading as fairly common active topology measurements.
First, NOISE generates nonces en masse and
embeds them as interface
identifiers (IID), \eg some lower 64 bits of an IPv6 address, resulting 
in a batch of ``nonced'' (nonce-laden) IPv6 addresses. For example:
{\tt 2001:db8::\textbf{dead:beef:f00d:cafe}}, where
the portion shown in bold is the nonce and the top 64-bit network
identifier is dictated by the prefix of NOISE's address block,
an IPv6 /36 prefix (having $2^{92}$ total addresses) 
never before used and dedicated solely to our experiments.
Specifically, NOISE nonces are a 64-bit value, 
\eg a monotonically increasing 64-bit counter, 
encrypted by the ChaCha20 stream cipher algorithm~\cite{bernstein2008chacha},
so the value is obscured and the nonce unpredictable.
If they were predictable or the encryption key compromised, an
adversary could craft and transmit valid nonces they did not actually observe
as the result of our transmissions,
misdirecting our analysis.

With our nonced IPv6 source addresses in hand, NOISE disseminates them
by running special traceroute-like campaigns. In regular traceroute, 
probe packets having monotonically increasing Time-to-Live values (TTL,
also known as ``Hop Limit'' in IPv6) 
are sent from the IP address of the source host to the targets 
of interest. 
NOISE emits traceroute-like, TTL-limited probes with crafted or {\em forged}
source addresses --- one-time-use, nonce-laden source addresses ---
rather than the host's usual address. 
Given an incredibly large range of possible nonce values, \ie
{\em 64 bits}, we can afford to place a unique nonce in {\em every} probe
packet that we emit. 
That is, every single probe sent, each having a
TTL value between 1 and a maximum of 32,
for example, has its own unique, nonced source address.
As with traditional traceoute, each probe's TTL limits the distance it
travels (measured in router hops), therefore limiting where, topologically,
this unique nonced-source address might be observed by a monitor.

In order to conduct our special traceroute campaigns, we make a 
modified version of yarrp\cite{imc16yarrp}, a tool 
that performs traceroutes in a pseudorandom, stateless way, allowing
for fast, Internet-scale measurements of topology. While the original yarrp, like
traditional traceroute, uses a single source IP address on the
localhost when emitting probe packets, our modified yarrp
uses a list of source addresses from a file prepared in
advance: one for each and every probe packet. In NOISE
yarrp campaigns, this list is comprised of millions of nonce-laden
source addresses that we generate. Running yarrp on a host dedicated to NOISE,
we trace from these nonced IPv6 source addresses 
to the approx. 15.2M target addresses used in prior 
work~\cite{Beverly:2018:IBS:3278532.3278559}, to the best 
of our knowledge, representing the largest IPv6 topology surveys to date. 

Naturally, the question arises as to how NOISE can collect
responses to our traceroute probes given that the source addresses
are forged and not those of the NOISE source host itself.
We do this by actively restricting forged source addresses to the $2^{92}$ addresses in NOISE's address block, a /36 prefix, which is under our complete control, and
by forwarding all packets destined to addresses within
that block to the NOISE source host.
This is accomplished in router configuration,
\ie using a ``static route'' on the source host's gateway router
under our control, which then propagates NOISE's prefix into routing tables 
globally, making NOISE an active Internet sink~\cite{yegneswaran2004design}.

\subsubsection{\bf Passive components}
While disseminating nonces via yarrp
campaigns and on a continuously ongoing basis afterwards, NOISE
listens to see who or what reacts with interest to our nonced source addresses.
NOISE captures all packets destined for its address block of
all possible nonced source addresses. So the arrival of an unexpected packet
destined for a nonced address, \eg
arrival of a packet that is not an ICMP Error message (from a router) but
having a source address that was {\em not} a yarrp target, may
represent interest by a monitor given that it must have observed the nonced
address to have used it subsequently as a destination.

NOISE also listens for DNS backscatter.
Experience and prior work~\cite{fukuda2018knocks} show
that a common reaction to unsolicited traffic or probes, 
\eg by firewalls, is to perform a
``reverse'' DNS query ({\tt ip6.arpa.} PTR query in IPv6) on the source
address.  NOISE assumes a reverse query for one of its unpredictable nonced source addresses
is an expression of ``curiosity,'' \eg on the part of some monitor that must have 
observed that nonced address.  Any reverse queries not carrying previously
NOISE-disseminated nonces are easily detected and disregarded.
NOISE captures all DNS traffic seen at its nameserver, exclusively dedicated
as authoritative for the project's forward and reverse zones, \eg 
reverse DNS nameserver for NOISE's IPv6 address block.

\subsubsection{\bf Overview} NOISE comprises the following seven elements and four types of detection (shown in bold), further detailed in Section~\ref{sec:results}:
$(1)$ a dedicated IPv6 address block for our nonced source addresses;
$(2)$ a source host running yarrp and tcpdump (enabling {\bf pcap} detection);
$(3)$ a gateway router under our control, routing outbound and inbound traffic;
$(4)$ two project domain names exclusively for use solely by the system;
$(5)$ a DNS nameserver running NSD on a VM, authoritative for both 
$(a)$ reverse queries in the dedicated address block and $(b)$ forward
queries in two NOISE project domains and running tcpdump 
(enabling {\bf rdns} and {\bf fdns} detection, respectively);
$(6)$ a web server running Apache2 hosting a publicly-accessible NOISE
web site having the domain name that is the PTR name for all nonced source
addresses and describing our yarrp use and how to opt-out; and
$(7)$ access to DNSDB, a passive DNS database, to determine when 
queries for NOISE-specific nonced addresses {\em or} domain names evidencing interest 
or monitoring were shared with this third party commercial database (enabling {\bf pdns} detection).

\subsubsection{\bf Limitations} An unavoidable limitation in any attempt to remotely detect
Internet surveillance is simply that surveillants cannot be detected until they act.
Secondly, while not a limitation of nonce-based inverse surveillance in general,
our NOISE implementation has limited vectors by which it disseminates
nonces: we place nonces only in IPv6 source addresses and
transmit them from just a single source host in a single address block.
This limits the paths or trajectories that nonces travel, constraining
detection results to a topologically limited set of networks and paths.

\subsubsection{\bf Ethical Considerations}
Our active measurement survey
has similar concerns to that of Beverly {\em et al.}~\cite{Beverly:2018:IBS:3278532.3278559}. We likewise
obtain permission from the network hosting our vantage to
perform the survey, limit traffic load by running yarrp at $1k$ packets
per second, avoid probing likely active end host addresses,
and provide a way for complainants to contact us, \eg via email
as advertised
both in an Internet registry for our address block and on a web site
(operated at the PTR name for probe source addresses).
In some experiments, where we emit probes that masquerade as WWW traffic, there is a potential risk that a surveillant or censor might incorrectly suspect that
destination hosts are participating in actual WWW transactions with NOISE, like in prior works~\cite{vandersloot2018quack,mcdonald2018403}.
We claim only that we have no interest nor reasonable way to map these addresses to individuals.

\subsubsection{\bf Experimental Evaluation}
Our evaluation consists of a series of experiments each comprising
16 contiguous yarrp campaigns, largely having differing sets of destination addresses,
mimicking those in Beverly {\em et al.}~\cite{Beverly:2018:IBS:3278532.3278559}.
Herein, we present and discuss the results of three such experiments 
we've chosen to highlight which show how parameters such as protocol, port number(s), and maximum TTL
may influence results.  Table~\ref{tab:expdata} shows, in bold, each experiment's
parameters and the name by which we will refer to it.
\input{exp-data-table_scaled}
In two experiments, NOISE probes masquerade
as QUIC traffic---UDP and port 443---with the hope that encrypted WWW traffic is
of interest to a monitor or surveillant.
The UDP:443c experiment sends UDP probes from nonced source addresses
to port 443 of our approx. 15.2M targets as if the probes
are from a QUIC {\em client}, having a random source port number.  
The UDP:443s experiment sends UDP probes with nonced source addresses 
from port 443 to random port numbers of the same targets, as if the probes are from a QUIC {\em server}.
In the third, the Ping experiment, we send ICMPv6 Echo Request probes to the
same targets, masquerading as a typical topological measurement survey.

\looseness=-1
Because yarrp, by design, randomly orders probes (with respect to TTL and
destination), a given trace to a destination, interspersed amongst millions,
may take hours or days.
It is only complete after each TTL value, 1 through the maximum, has been transmitted. Throughout our
series of experiments, we decreased max. TTL from 32 (initially)
to 24 (when we saw 32 was unnecessarily high) to 16 (employing
yarrp's ``fill mode,''~\cite{Beverly:2018:IBS:3278532.3278559}
which goes beyond 16 as long as responses continue to be received).

%% file: exp-data-table_scaled.tex

\begin{table*}
  \centering
  \scriptsize
  \caption{Experiment Parameters and Resulting Data Characteristics}
  \label{tab:expdata}
  \begin{tabular}{rllrrrrrr}
    Exp. Name & Description & Maximum TTL & Dates, {2019} & Traces Performed & Dest. Addresses & Nonces Xmitted & Packets Captured\\
    \hline
    {\bfseries UDP:443c} & {\bf UDP probes sent TO port 443 } & {\bf 32 } & Jan  4 --10 & 15.2M & 12.4M & 486.9M & $>$ 652M \\

    {\bfseries UDP:443s} & {\bf UDP probes sent FROM port 443 } & {\bf 24 } & Apr 10 --14 & 15.2M & 12.4M & 365.2M & $>$ 495M \\

    {\bfseries Ping} & {\bf ICMPv6 Echo Request probes} & {\bf 16 +} & Apr 15 --18 & 15.2M & 12.4M & 311.5M & $>$ 396M \\
  \end{tabular}
\end{table*}

%% file: data.tex
\section{Data}
\label{sec:data}

\looseness=-1
Data resulting from performing the three experiments is summarized
in Table~\ref{tab:expdata}.
Note that because we use a unique nonce in each trace
packet transmitted (``xmitted''), the ``Nonces Xmitted'' value is
also the number of packets transmitted across all traces to the
destinations (addresses).
The 15.2M trace destinations (aka target addresses, of which 12.4M are unique) 
were chosen as in Beverly et al.\cite{Beverly:2018:IBS:3278532.3278559}, which
is the largest IPv6 trace survey of which we are aware.
We refer the interested reader to that paper for myriad target selection
details we have not repeated here.

In total, we emitted approx. one billion
unique nonce-laden packets destined for IPv6 networks worldwide. 
Recall that on our trace source host
and DNS server, we capture all transmitted probes and all received
packets of interest. We represent our voluminous trace and packet data in a
graph database, which proved valuable in analyzing and presenting
the following results.

%% file: results.tex
\section{Results}
\label{sec:results}

From the overview in Section \ref{sec:method}, recall that NOISE has four detection
types: {\bf rdns} (reverse DNS), {\bf pcap} (packet capture), 
{\bf pdns} (passive DNS), and {\bf fdns} (forward DNS). 
Each of these represents a reaction in response to
a probe packet, having a nonced source address, being received or observed
some distance from the source, \ie remotely monitored.
An rdns detection means a nonced address was reverse looked-up. 
A pcap detection means an unexpected packet was received, destined for a
nonced source address in a probe we sent (\eg ping of a nonced address).
A pdns detection means a nonced address was found in a
reverse DNS query recorded in DNSDB, a commercial passive database.
An fdns detection means a forward DNS type A or AAAA query was performed
on the PTR name provided by the NOISE authoritative DNS server as an answer
to a prior reverse DNS query on a nonced address. 
With this in mind, we present a ``macroscopic'' view of our results 
followed by instances of detailed, ``microscopic'' views and validation of our results.

\subsection{Macroscopic View} Across three experiments,
NOISE detects monitoring more than 200k times,
ostensibly in 268 networks, for probes destined for 437 networks.
When monitoring was detected, it resulted from approx. 25k of the traces
having target destinations in a total of 55 countries. The top five were the 
United States (9,099), Germany (3,805), Brazil (2,726), 
Switzerland (1,850), and the United Kingdom (1,502). 
(Our NOISE source host/vantage point is in the United States.)

\begin{figure}
  \includegraphics[width=\columnwidth]{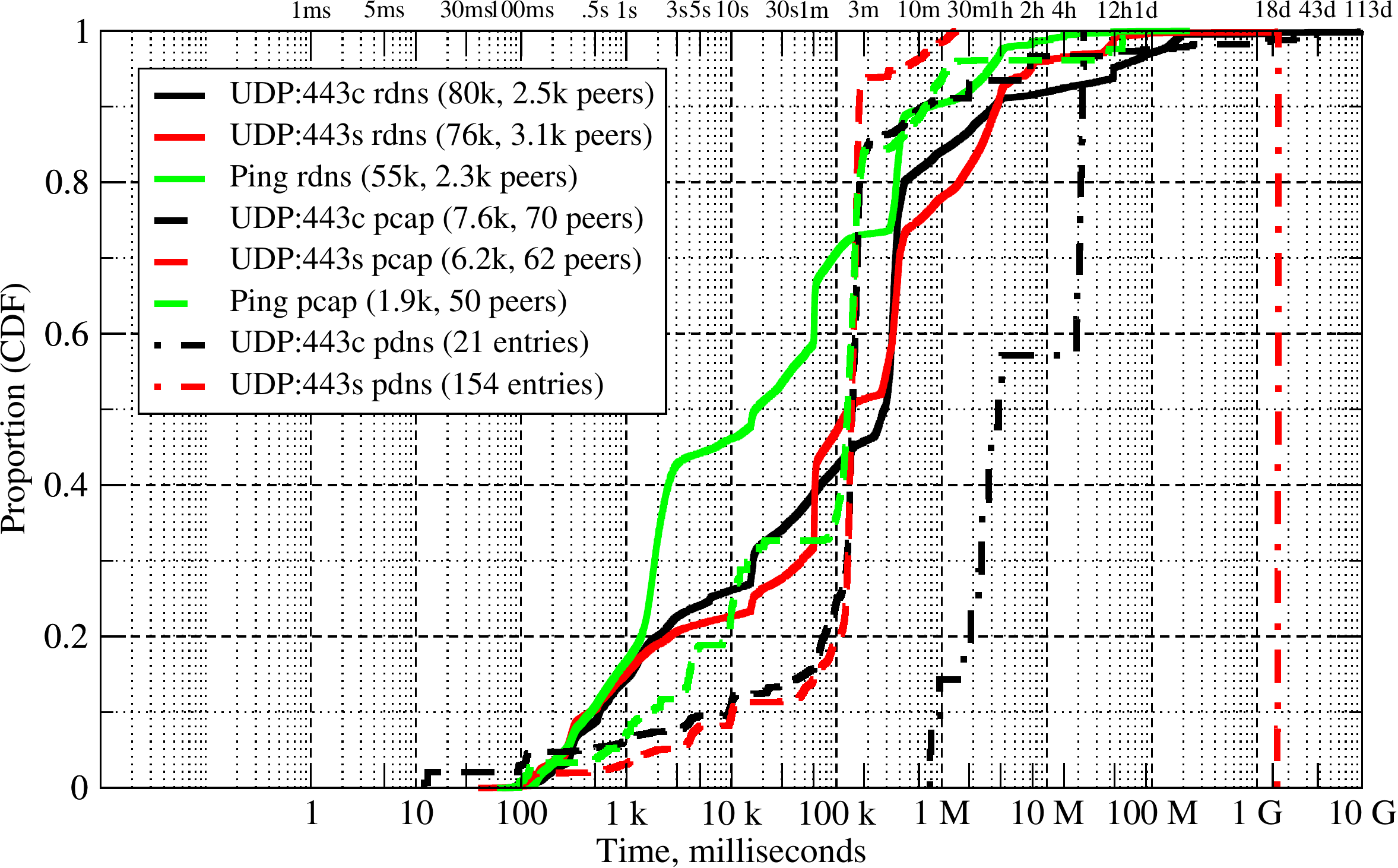}
  \caption{Times to detection of nonce propagation by type, per experiment.
  \label{fig:correct-temporal}}
\end{figure}

In \fig~\ref{fig:correct-temporal}, we show the counts of monitoring
by detection type and the amount of time to each type of detection.
The number of detections by type is in parentheses, along with 
the number of peer hosts (unique remote addresses) that were the
source of the reaction, where applicable.
There was a total of approx. 247k reactions across 
all three experiments, and these occurred in approx. 25k traces to 
approx. 17.4k targets. The rdns detections were the most prevalent by far. 
Most detections occurred within 24 hours. However, there are some 
outliers: an rdns detection 113 days later in our 
UDP:443c experiment and a pcap detection that happened 42 days later. 
We go into 
detail about the pdns detections from experiment UDP:443s that happened 
18 days later in Section~\ref{sec:validation}. 

\begin{figure}[h]
  \includegraphics[width=\columnwidth]{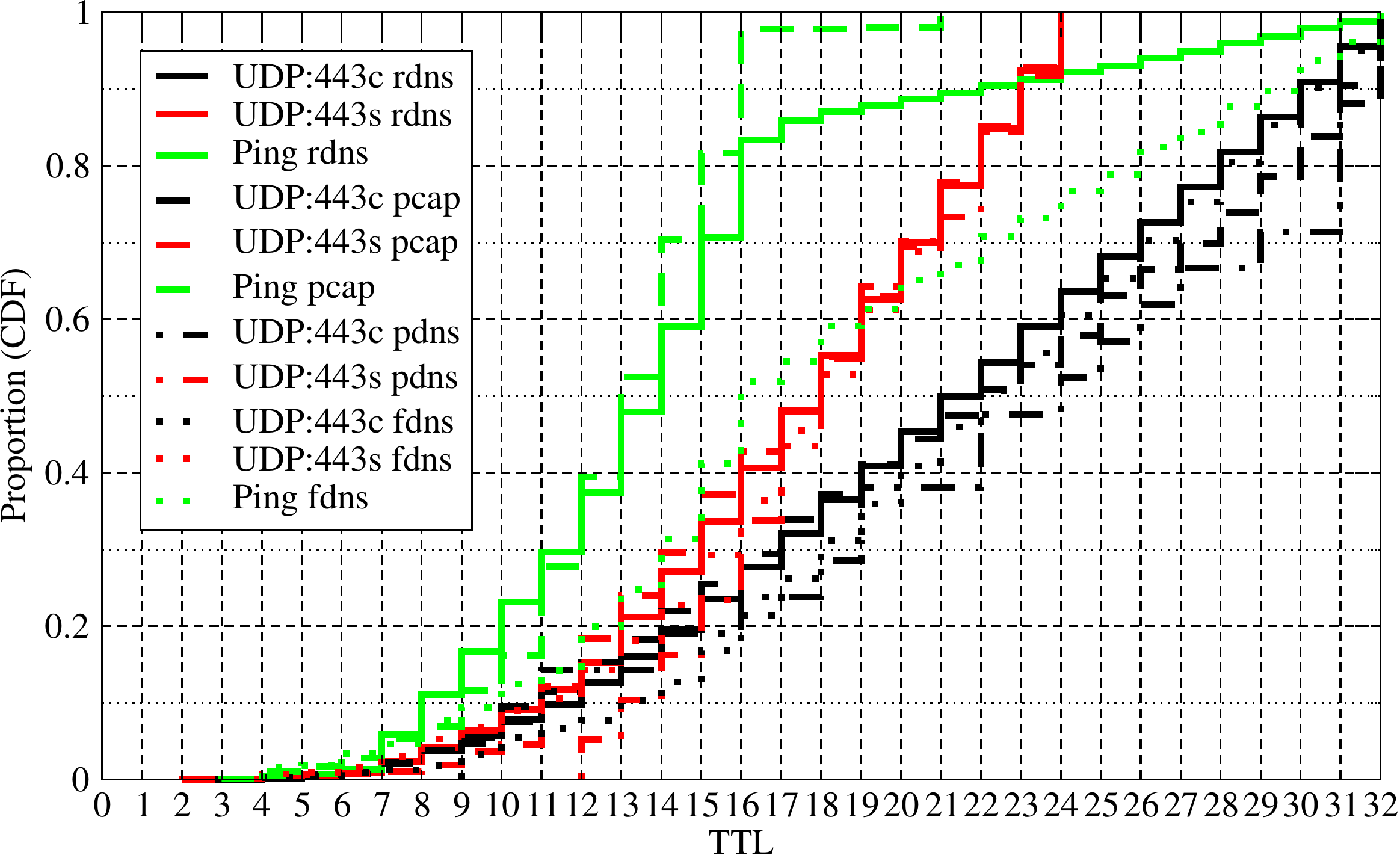}
  \caption{Probe TTLs for detected nonce propagation by type, per experiment.
  \label{fig:spatial}}
\end{figure}

We also examine which TTL values were associated with the probes
generating the most reactions or detections.
\fig~\ref{fig:spatial} summarizes our results for rdns, 
pcap, pdns, and fdns reactions. 
Reactions to probes having very low TTLs are especially interesting
(\ie TTLs toward the left side of the graph) 
because these likely represent monitoring in the middle, rather than
at the edge, in target networks.
In Section~\ref{sec:validation}, we present details for the
leftmost instance, having TTL of only 2, where eavesdropping was detected.

\input{observationASN-not-in-dstASN}
\input{rdns-networks-table}
Of special interest is monitoring 
by surveillants who are ostensibly not in the target 
destination network of a trace but rather somewhere in the middle.
We report detection counts in Table~\ref{tab:observationASNs} where
the reacting remote peer address is identified by origin
ASN, \ie the Autonomous System Number that
originates a route via the global BGP (Border Gateway Protocol) covering that remote peer address.
With rdns detections, we often see
the origin ASN for the source of a reverse query (for a nonced address)
is not the origin ASN for the trace's target/destination address. Indeed,
Table~\ref{tab:rdns-networks} shows the top 10 origin ASNs of the client
remote peer addresses that performed the reverse DNS queries; note that
they include popular, public recursive DNS service providers.
This strongly suggests 
that monitors and surveillants propagate their queries to third parties, 
resulting in some traffic information being disseminated to them.

The pcap detections involving
remote source addresses having different origin ASN than that of the trace
target destination address are also suspicious. The top ASNs associated
with pcap reactions (not shown, to preserve those networks' anonymity)
have as many as 40 unique source addresses
sending packets to nonced addresses.

\begin{figure}[h]
  \includegraphics[width=3.2in]{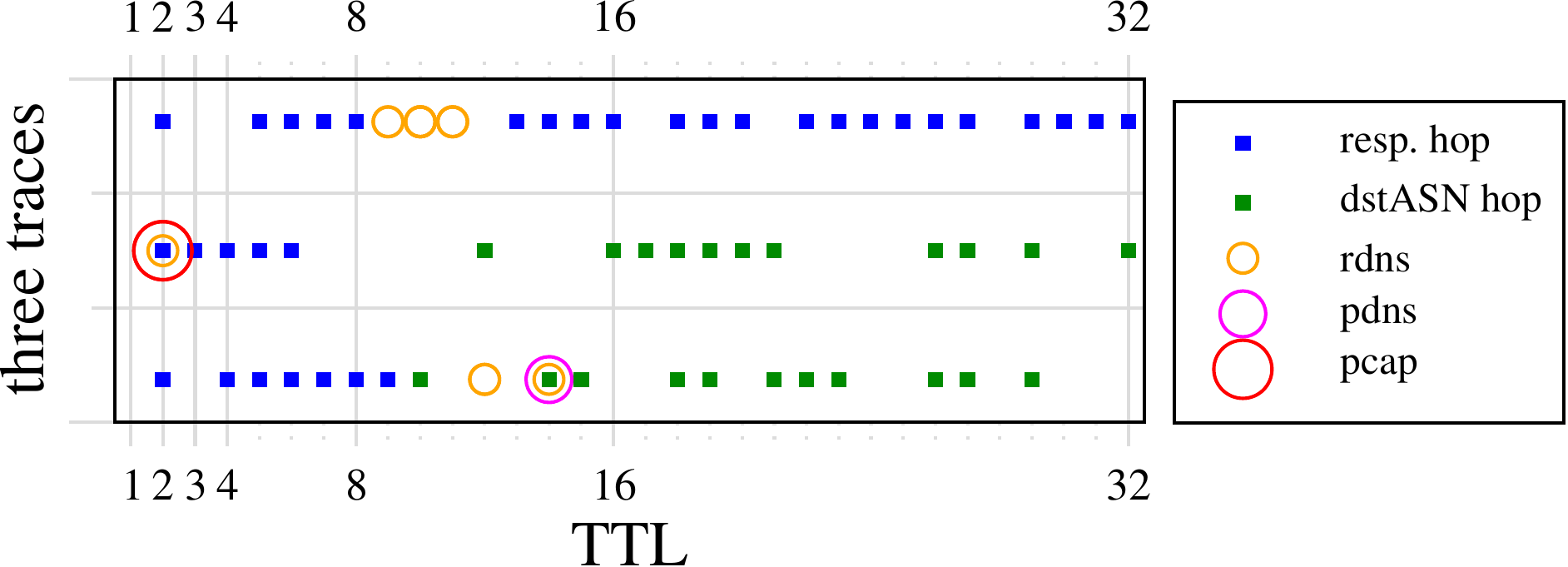}
  \caption{A visualization of three UDP:443c experiment traces (to three targets) showing evidence of monitoring. 
    The horizontal axis is spatial by TTL, \ie the distance increases from left to right.
  \label{fig:exemplar}}
\end{figure}

\looseness=-1
In more detail next, we visualize trace-level results in
aggregate to see how monitoring detection varies by experiment parameters
and varies across
the targets' address space and origin ASNs.
To introduce this visualization
technique, \fig~\ref{fig:exemplar} shows three traces from 
experiment UDP:443c that contain rdns, pcap, and pdns detections. 
In this visualization, the traces are represented from left to right, from 
TTLs 1-32. 
The solid blue squares represent responsive hops. 
The solid green squares represent responsive hops that were in the target 
destination ASN of the trace. 
The small orange circles (rdns) indicate that a 
reverse lookup was performed on the nonced address of the probe we sent with that 
TTL.
The bigger magenta circles (pdns) indicate that we found the nonced 
address of the probe we sent having that TTL subsequently present in DNSDB.
Finally, the big red circles (pcap) indicate that
a packet was unexpectedly received destined for the nonced address of the probe we 
sent having that TTL.
Note that if we detect a reaction to a nonce-laden packet sent with
TTL of 9, for example, it does {\em not} mean that a monitor or surveillant was at 
exactly hop 9. It means that monitoring ostensibly occurred within 9 hops 
along the path to the target because hops 1-8 also had the opportunity
to observe the probe packet with a TTL of 9. The lower the TTL value
associated with detection, the lower the {\em upper limit} on topological distance
to the monitor and typically, the more constrained the observer's possible location.

In the top trace, we see that our probes might
not have reached the target's origin ASN because it has only blue
responsive hops, not green. (It's possible the probes
were responded to, and possibly filtered, by some network upstream
from the destination, one that is possibly affiliated with it.)
In the middle trace, we see monitoring detected by 
rdns and pcap methods involving the nonced address for a probe sent with a TTL of 2.
In the bottom trace we see evidence of  a monitor
based on rdns and pdns detection. Notice that these are probes 
that reached the target's origin ASN.
(Section~\ref{sec:validation} has details validating monitor detection
for these particular traces.)

\begin{figure}
  \includegraphics[width=\columnwidth]{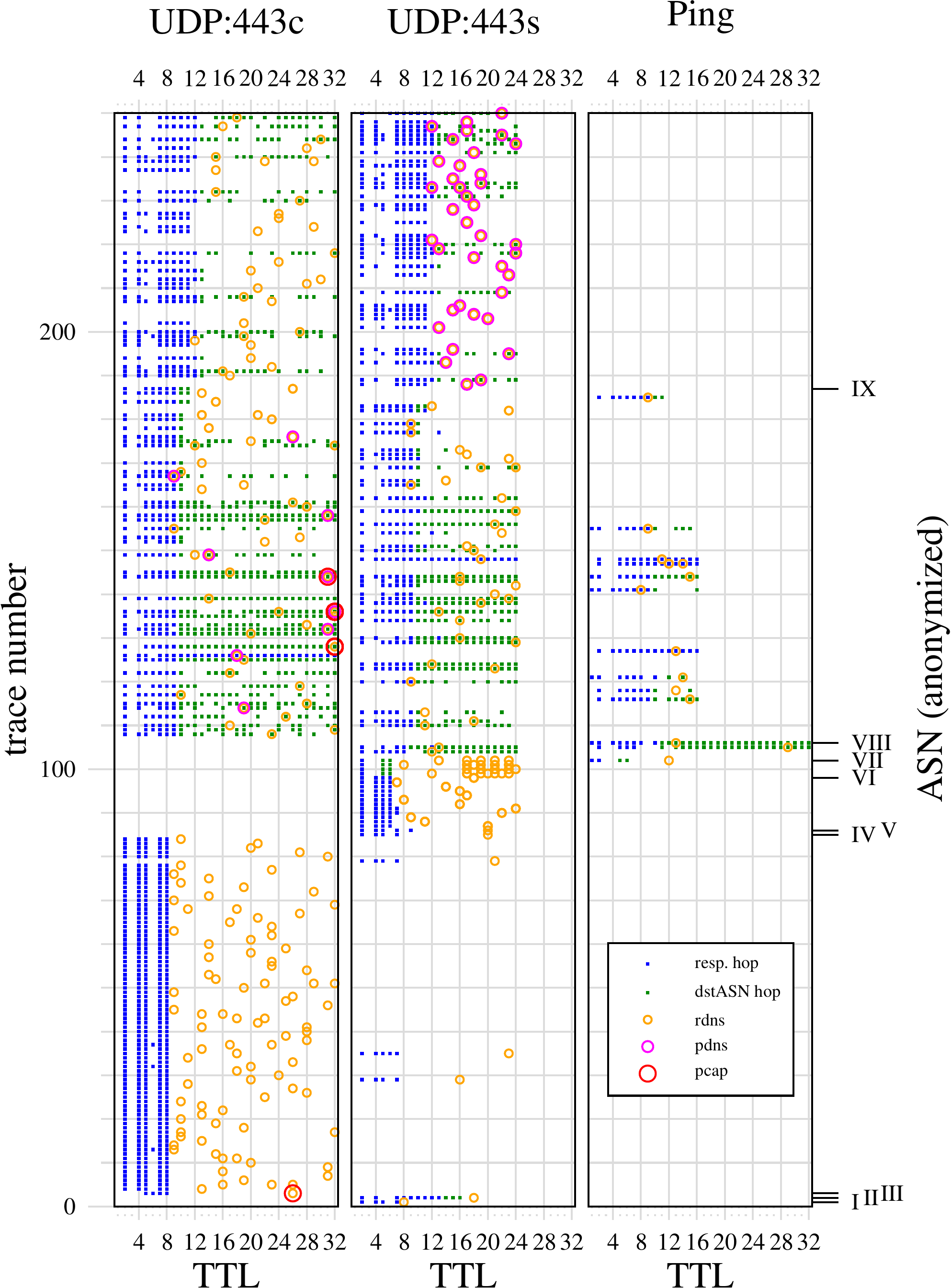}
  \caption{A visualization of nearly 250 traces (to unique targets) across each of our three experiments, having evidence of monitoring in at least one experiment.
  \label{fig:500}}
\end{figure}

With this knowledge, we can move on to looking at our similar, but bigger,
visualization in \fig~\ref{fig:500}. This displays trace 
data for nearly 250 of the approx. 25k traces where monitoring was
detected.  Because we trace to the 
same targets in each experiment, we line-up the traces having the same
target (horizontally), exposing how monitor behaviors change either over
time or due to differing probe types per experiment.
The traces are arranged on the vertical axis by destination ASN and in 
order of target address within each ASN,
and we display divisions between targets' origin
ASNs on the rightmost label of the vertical axis.

First, comparing experiments across \fig~\ref{fig:500},
note that there are routers that 
respond to UDP probes to or from port 443 that did not respond to 
our ICMPv6 ping probes, and the same can be said of monitors or surveillants.
The differences between experiments UDP:443s and Ping are 
particularly interesting given that they were run back-to-back. 
For example, the topmost ASN ignored our ICMPv6 probes but was very 
responsive to, or arguably {\em interested in,} probes having source port 443.
This reveals that monitoring practices and policies clearly differ
by ASN. It also indicates that probe protocols matter when designing 
systems to detect monitoring.
While space limitations preclude describing all the phenomena 
evident in \fig~\ref{fig:500}, we claim it demonstrates
NOISE's power to identify monitoring and active response practices
and associate them with specific networks' address blocks, albeit
shown anonymously, here.

\subsection{Microscopic View \& Validation}
\label{sec:validation}
Given the myriad instances of monitoring we've detected
in the modest set of experiments presented, we next attempt to
validate or verify a subset of the results by gathering
ground-truth for instances of NOISE-detected monitoring: 
curious DNS queries, sharing DNS data,
and eavesdropping or ``monitoring in the middle.''
Each of these anecdotes entails
monitoring by, or sharing resulting information with, {\em third
parties} who are neither the source nor target networks of
NOISE probes.

\subsubsection{\bf Curious Queries}

Table~\ref{tab:asiancountry} shows an abbreviated timeline of events
for a trace in which NOISE received reverse DNS (rdns) queries,
indicating monitoring, \ie a reverse lookup
was performed on the nonced source address of a probe packet. The
lines in bold represent evidence of monitoring within 10 hops
of the NOISE source host.  At 9m 7s and seconds
thereafter, NOISE captures reverse DNS queries for the nonced 
address sent with hop limit of 10.  The source address of these query 
packets belongs to the network containing 
the trace's target address.

\input{asian-country-anecdote-table}

This trace is also the top trace in \fig~\ref{fig:exemplar},
where we see the NOISE observations (orange circles) occur at hops 9,
10, and 11, although we did not receive an ICMPv6 hop limit exceeded error message from
routers at these hops.  Note that no responsive hops are in a network
associated with the target (that is, only blue squares). Here
a traditional traceroute, without nonces, does not show whether or not probes reach the
destination network.  However, NOISE's rdns detection provides evidence that
trace probes' nonces {\em did} reach the destination network of the target
because the recursive DNS
queries had source addresses within the target's destination network.
This demonstrates how NOISE's modified yarrp sometimes improves reachability
measurements over those
performed with either yarrp or other traceroute tools.

Furthermore (but not shown), using NOISE's detailed packet capture logs at its
authoritative DNS service (NSD), we find that this network's
DNS server(s) subsequently queried our NOISE DNS server by the {\em
name} we provided as an answer to the PTR query, \eg {\tt
{\em something}.noise.example.com}. This exposes a network vulnerability
that could be abused.  Once such behavior is known, an adversary
could remotely cause the institution to query
an arbitrary domain name of the adversary's choosing merely by
replying with that name in response to a PTR query (that it can
elicit using NOISE's nonced address probing technique). For example,
this can be abused for $(a)$ misdirection (of network forensics
investigation), $(b)$ adversely affecting caching and performance of
the institution's DNS service, and $(c)$ the institution's unwitting
participation in DNS amplification attacks.

\subsubsection{\bf Sharing Passive DNS Data}

\input{university-anecdote-table}

Table~\ref{tab:university} shows an abbreviated timeline of events
for a trace in which NOISE found that queries involving its nonced addresses
were somehow conveyed to DNSDB.  This trace,
performed in April 2019, was from the NOISE source host located
within a commercial datacenter in the U.S. to a target at a
U.S. university.  The two lines in bold represent strong evidence
of monitoring within 14 hops of the NOISE source host.  At 1h
47s in the trace, NOISE transmits a UDP probe having a nonced
address and port 443 as its source and a hop limit of 14. At
4h 44m, the trace is complete.  At 18d 5h---18 {\em days} later---NOISE 
captures a reverse DNS query for the nonced address sent
with hop limit of 14.  The source address of this query packet belongs to
the university's network containing the trace's target address.
At 18d 6h---an hour later---this nonced address appears in the
third-party passive DNS database.

A similar trace, performed in January 2019, destined for a
different U.S. university, is the bottom trace in
\fig~\ref{fig:exemplar}. There you can see at hop 14 that NOISE
has both rdns (reverse DNS, orange circle) and pdns (passive DNS,
magenta circle) observations.
Across all three experiments, we found that some of our nonced
addresses ultimately appeared in DNSDB
when the NOISE probe packets reached either of two unrelated
U.S. universities.  That is, their PTR queries and responses were captured in
this database somehow by that third-party's network of monitored
DNS servers.  This means that NOISE can identify some recursive
nameservers that have a third party's monitor installed
and are monitoring DNS queries and answers and
transmitting them to the company.

To get to the truth, we had personal conversations with expert
operations personnel at each of the two universities, agreeing to maintain
their anonymity.  The first university, associated with
pdns detections on traces in January 2019, reported that
network operations collected data from border routers via flow
export~\cite{hofstede2014flow} to support network troubleshooting and
forensics.  This data was then post-processed using custom scripts
that sometimes perform DNS reverse lookups~\cite{netJune2019}.
Furthermore, they reported that the university had been running
passive DNS query monitoring software in the university's primary
DNS server infrastructure in January 2019 and years prior.
A security officer also reported that their operations systems
rely on this DNS infrastructure for recursive queries, and thus, DNS
queries in incident handling may very well be subject to passive DNS
monitoring~\cite{secJune2019}.  Coincidentally, they also reported
that university technical personnel decided not to reinstall this
passive DNS software when the DNS server infrastructure was upgraded
prior to April 2019.  Our results coincide with this: NOISE
no longer had pdns detections associated with this university as
of April 2019.

\looseness=-1
Similarly, regarding the second university, we validated our
results by personal, anonymous interview. For traces like the one 
in Table~\ref{tab:university}, we learned that this
university's network team records {\em all} traffic meta-data
via flow export, not just unsolicited traffic such as our
probes~\cite{secJune2019}.  Furthermore, the reported purpose of
the data collection was to support incident handling and network
troubleshooting~\cite{dirJune2019}. With respect to the 18 days
passing between the time the NOISE probe was received and the time
the reverse DNS lookup was performed, they hypothesized that the
subsequent lookup was due to an actual incident investigation,
likely attended to by an analyst, \ie some manual effort.
Their hypothesis is that the probe matched an automatic detection
rule targeting threats which would cause it to garner additional
attention. They did not share their log retention policy, but it
is clear from our results that logs were retained for at least 18 days at the time.
They also reported that the university indeed operates passive DNS 
monitoring as is a prescribed best practice amongst some higher-education
institutions~\cite{renIsacPdns}.

\fig~\ref{fig:500} includes for both these universities pdns
observations in context (magenta circles). These two universities' 
ASNs are arranged above and below the tick mark IX, topmost
in the vertical axis labels on the right. First, considering the traces 
above that 
tick mark (trace numbers 187-250), we see that in the first
column (UDP:443c, January 2019) of traces destined for UDP port 443, there are
many rdns detections (orange circles) but no pdns
detections (magenta circles).
In the second column (UDP:443s, April 2019) of traces, however,
from UDP port 443 and destined for pseudorandom ports, there are
many pdns observations (magenta circles). Lastly for this ASN, in the third column
(Ping, April 2019) of traces, ICMPv6 echo requests show no detections.  The
Ping experiment's traffic is treated differently, either in monitoring or
in reactions, \eg automated or manual analysis.
In \fig~\ref{fig:500}, consider the traces in the other
university's ASN, below tick mark IX, above tick mark VIII
(trace numbers 106-186).
Here we see occasional pdns detections
in the first column (UDP:443c, January 2019), evidencing that
institution's participating in passive DNS monitoring at the
time of that survey experiment. However, these observations do not
appear in the latter columns (UDP:443s and Ping, April 2019),
coinciding with the university personnel reporting they were no
longer performing the passive DNS monitoring.

\looseness=-1
Overall, this remote detection of passive DNS could be used and 
abused in some ways. Because the NOISE technique can remotely detect
data collection and monitors, potential adversaries could use NOISE
to classify
institutions' networks and attack surfaces. For instance, one might
attack collection infrastructure by employing the NOISE technique 
to cause certain institutions to perform many queries. 
This can pollute 
the passive DNS database with either misleading or superfluous
information, causing confusion or operational problems that
could be detrimental to security investigations. The ability
to remotely cause a network to perform DNS lookups and cause
the resulting names and addresses to be stored, indefinitely,
in passive DNS databases also presents subsequent security
or privacy issues, \eg as addresses or names in those
long-lived records become encumbered by reputation or become
targets~\cite{gasser2018clusters,Beverly:2018:IBS:3278532.3278559}
as they pass from one third party to the next.

\subsubsection{\bf Eavesdropping}

Table~\ref{tab:mitm} shows an abbreviated timeline of events for a
trace in which NOISE detected monitoring amid the trace path from source to
destination ASN, \ie a monitor in the middle.
This trace is the middle trace in \fig~\ref{fig:exemplar},
where we see the NOISE detections (circles) occur at hop 2 (a blue square),
well before the trace reaches the destination ASN by hop 12 (green squares).
This trace, performed in January 2019, was from the NOISE source
host, located within a commercial datacenter in the U.S., to a
U.S. target thousands of miles away.  The three lines in bold
represent strong evidence of eavesdropping within two hops of the
NOISE source host.  First, to  start the trace, NOISE transmits a
UDP probe having a nonced address as its source and a hop limit of 2,
destined for port 443 of the target.  At 9m 58s, NOISE unexpectedly captures a TCP
SYN packet destined for port 80 and the nonced address used as source
of the hop-limit=2 probe.  At 10m 25s, NOISE captures a similar
TCP SYN packet destined for port 443 and this same nonced address.
Both these TCP SYN packets have an unfamiliar source address belonging
to a popular cloud host provider, different than the datacenter.
At 10m 43s, NOISE captures a
reverse DNS query for this same nonced address.  The source address
of this query belongs to a public recursive DNS provider's network.

\input{mitm-anecdote-table}

Notice that the second line of Table~\ref{tab:mitm}, at 0.0005s,
shows that NOISE captured a router hop response for hop limit
2 after only a fraction of a millisecond, \ie an ICMPv6
hop limit exceeded error message was received, destined for the nonced
source address. This means that the nonce was first propagated
by a router {\em before} the eavesdropping detection events (bold lines),
and thus, it is possible that an eavesdropper gleaned the
nonced address from the ICMPv6 error message packet rather than from the
NOISE probe packet. However, we further find that that ICMP error message
packet itself had a hop limit of 31 when it arrived at the NOISE
source host. This suggests its initial hop limit was 32 because
common host implementations are known to use 32, 64, or 255. If
so, the ICMPv6 error response traversed just one router,
suggesting it originated at a router at hop 2 and that its
return path has a maximum length of two hops.
Given {\em (a)} a rule-of-thumb that 1ms round-trip-time
(RTT) represents a maximum distance of approx. 
100km
and {\em (b)} the RTT for a router at hop 2 is only
0.5 milliseconds, we can reasonably assume that this router is within
50km of the NOISE source host. Consequently, hop 2 is definitely ``in
the middle'' given that the trace target is thousands of miles away,
bolstering a conclusion that this evidences eavesdropping.

With the help
of the host network's architect, we have eliminated the
possibility that the eavesdropping is being performed in
the NOISE host network itself, \ie within the first hop~\cite{netJune2019}.  All indications are that the eavesdropping
is on a link between the host's gateway router (belonging to
the host network) and the next-hop router, operated by an ISP with
which the host network peers.

\looseness=-1
Using NOISE's detailed
logs, we further find that a little more than 10 minutes passed 
between dissemination of the nonce in the probe packet (and
presumed eavesdropping) and propagation of that nonce back to our
authoritative name server by third-party reverse query.
Thus, NOISE can detect some monitoring (\eg 
packet or flow capture or logging), can then isolate its candidate location
in the router-level Internet topology (path), and can ultimately show that information
about traffic is being shared with third parties.
Across all three experiments, we find a total of 459 traces having
one of these suspicious TCP SYN connection attempts from the
given cloud host provider's ASN to 459 distinct NOISE nonced addresses.
The distribution of hop limits associated with those nonced
source addresses (not shown) suggests uniform packet sampling,
irrespective of hop limit.

Once NOISE identified the cloud host provider's source IP address used
in these curious TCP SYN connection attempts, we searched for independent
records of this suspicious behavior in the security community.
This search yielded two pieces of evidence involving the
source address's /64 prefix.
First, a large Content Delivery Network's transaction logs were searched for the /64 prefix in question as a WWW client.  We find that it is
regularly the source of myriad successful connections to WWW infrastructure~\cite{archJune2019}.
Second, we find an independent report online 
(late 2018, not included here to maintain privacy) 
that it has been the source of connection attempts to IPv6 {\em temporary privacy addresses} 
in another country, consistent with the monitor’s being in a privileged location, 
such as an Internet exchange.

%% file: observationASN-not-in-dstASN.tex
\begin{table}
  \centering
  \scriptsize
  \caption{Detection counts where remote peer host's origin ASN differs from that of Trace Target Destination}
  \label{tab:observationASNs}
  \begin{tabular}{rcrrr}
    Exp. Name & Detection & \# Reactions from & Total \# & \% \\
    & Type & \multicolumn{1}{r}{Diff. DstASN} & \multicolumn{1}{r}{Reactions} & \\
    \hline
    \multirow{3}{*}{UDP:443c} & 
    	    {\bfseries rdns} & 34,306 & 79,552 & 43.12 \\
        &{\bfseries pcap} & 2,003 & 7,625 & 26.27 \\
        &{\bfseries pdns} & n/a & 21 & n/a \\
    \hline
    \multirow{3}{*}{UDP:443s} &
        {\bfseries rdns} & 28,615 & 76,154 & 37.58 \\
        &{\bfseries pcap} & 1,191 & 6,237 & 19.10 \\
        &{\bfseries pdns} & n/a & 154 & n/a \\
    \hline
    \multirow{3}{*}{Ping} &
        {\bfseries rdns} & 29,812 & 54,663 & 54.54 \\
        &{\bfseries pcap} & 248 & 1,869 & 13.27 \\
        &{\bfseries pdns} & n/a & 0 & n/a \\
  \end{tabular}
\end{table}

%% file: rdns-networks-table.tex
\begin{table}[t]
\centering
\scriptsize
\caption{Top 10 origin ASNs for remote addresses
performing PTR queries on nonced addresses (rdns), in one experiment}
\begin{tabular}{rrrl}
Exp. Name & \# NS addrs & ASN & AS Name\\
\hline
\multirow{10}{*}{UDP:443c} &
1,277 & 15169 &  Google LLC\\
&175 & 13335 &  Cloudflare, Inc.\\
&139 & 36692 &  OpenDNS, LLC\\
&85 & 3356  &  Level 3 Parent, LLC\\
&83 & 8075  &  Microsoft Corp.\\
&63 & 9355  &  NICT\\
&62 & 24940 &  HETZNER-AS\\
&53 & 3462  &  HINET Data Comm. Business Group\\
&38 & 4782  &  GSNET Data Comm. Business Group\\
&34 & 42    & WoodyNet\\
\end{tabular}
\label{tab:rdns-networks}
\end{table}

%% file: asian-country-anecdote-table.tex
\begin{table}
  \centering
  \scriptsize
  \caption{Events that occurred during trace to an Asian network in Experiment UDP:443c}
  \label{tab:asiancountry}
  \begin{tabular}{rlr}
    Delta time & Event & ProbeTTL \\
    \hline
    0s & tr probe sent to target & 26 \\
    0.24s & tr hop response & 26 \\
    8m 56s & tr probe sent to target & 10 \\

    {\bfseries 9m 7s} & {\bfseries RDNS query on {\em noncedAddr}} & \\ 
    &                      {\bfseries by target's network} & {\bfseries 10} \\

    {\bfseries 9m 10s} & {\bfseries RDNS query on {\em noncedAddr}} & \\ 
    &                      {\bfseries by target's network} & {\bfseries 10} \\

    3h 6m & tr probe sent to target & 14 \\
    3h 6m & tr hop response & 14 \\

    3h 38m & tr probe sent to target & 32 \\
    3h 38m & tr hop response & 32 \\

    {\vdots} & {\vdots} & {\vdots} \\
    1d 15h & last tr probe sent to target & 29 \\
    1d 15h & tr hop response & 29 \\
  \end{tabular}
\end{table}

%% file: university-anecdote-table.tex
\begin{table}
  \centering
  \scriptsize
  \caption{Events that occurred during and after trace to a university in Experiment UDP:443s}
  \label{tab:university}
  \begin{tabular}{rlr}
    Delta time & Event & ProbeTTL \\
    \hline
    0s & tr probe sent to target & 15 \\
    2m 32s & tr probe sent to target & 17 \\
    16m 14s & tr probe sent to target & 7 \\
    16m 14s & tr hop response & 7 \\
    
    {\vdots} & {\vdots} & {\vdots} \\
    1h 47s & tr probe sent to target & 14 \\
    {\vdots} & {\vdots} & {\vdots} \\

    4h 44m & last tr probe sent to target & 4 \\
    4h 44m & tr hop response & 4 \\

    {\bfseries 18d 5h} & {\bfseries RDNS query on {\em noncedAddr}} & \\ 
    &                      {\bfseries by university} & {\bfseries 14} \\

    {\bfseries 18d 6h} & {\bfseries {\em noncedAddr} appears in} & \\ 
    &                      {\bfseries passive DNS database} & {\bfseries 14} \\
  \end{tabular}
\end{table}

%% file: mitm-anecdote-table.tex
\begin{table}
  \centering
  \scriptsize
  \caption{Events that occurred during trace detecting Eavesdropping in Experiment UDP:443c}
  \label{tab:mitm}
  \begin{tabular}{rlr}
    Delta time & Event & ProbeTTL \\
    \hline
    0s & tr probe sent to target & 2 \\
    0.0005s & tr hop response & 2 \\
    {\bfseries 9m 58s} & {\bfseries TCP SYN :20 $\rightarrow$ {\em noncedAddr}:80} & \\ 
    &                    {\bfseries by cloud Provider} & {\bfseries 2} \\
    {\bfseries 10m 25s} & {\bfseries TCP SYN :20 $\rightarrow$ {\em noncedAddr}:443} & \\
    &                     {\bfseries by cloud Provider} & {\bfseries 2} \\
	{\bfseries 10m 43s} & {\bfseries RDNS query on {\em noncedAddr}}  & \\
	&                     {\bfseries by cloud DNS Provider} & {\bfseries 2} \\
    22m 26s & tr probe sent to target & 24 \\
    {\vdots} & {\vdots} & {\vdots} \\
    11h 51m & last tr probe sent to target & 15 \\

  \end{tabular}
\end{table}

%% file: contributions.tex
\section*{Conclusion}

Motivated by concerns of our own and of the community about
pervasive, systematic surveillance of Internet traffic,
we develop NOISE: an inverse surveillance method and system.
We've evaluated and reported its effectiveness,
but this is only a start. In general, the method is
not limited to only nonce-laden (IPv6) transport identifiers,
nor limited to traffic synthesized for active measurements.
We envision broad application in concert with, and
literally {\em within}, everyday Internet traffic and applications.
With NOISE implemented pervasively, \eg in the WWW,
monitors would have no choice but to observe nonce-laden traffic,
improving detection of surveillants whenever they act on their observations.

%% file: acknowledgment.tex
We appreciate the significant help on this work from these 
colleagues and coworkers:
Niels Bakker,
Arthur Berger,
Robert Beverly,
Aaron Block,
David Choffnes,
David Duff,
Jared Mauch,
Suzanne Pan,
Philipp Richter,
Kyle Rose,
Steven Schecter,
Chris Schill,
Jon Thompson,
and Rick Weber.
This work utilizes tshark, GNU parallel, and Neo4j.~\cite{combstshark,tange2011gnu,neo4j}